\documentclass[twocolumn,showpacs,preprintnumbers,amsmath,amssymb,prl,superscriptaddress]{revtex4}

\usepackage{graphicx}
\usepackage{dcolumn}
\usepackage{bm}

\begin{document}


\title{Dynamic nuclear polarization using a single pair of electrons }

\author{S.Foletti}
\affiliation{Department of Physics, Harvard University, Cambridge, MA 02138, USA}
\affiliation{Dept. of Condensed Matter Physics, Weizmann Institute of Science, Rehovot 76100, Israel}

\author{J.Martin}
\affiliation{Department of Physics, Harvard University, Cambridge, MA 02138, USA}
\affiliation{Dept. of Condensed Matter Physics, Weizmann Institute of Science, Rehovot 76100, Israel}

\author{M.Dolev}
\affiliation{Dept. of Condensed Matter Physics, Weizmann Institute of Science, Rehovot 76100, Israel}

\author{D.Mahalu}
\affiliation{Dept. of Condensed Matter Physics, Weizmann Institute of Science, Rehovot 76100, Israel}

\author{V.Umansky}
\affiliation{Dept. of Condensed Matter Physics, Weizmann Institute of Science, Rehovot 76100, Israel}

\author{A.Yacoby}
\affiliation{Department of Physics, Harvard University, Cambridge, MA 02138, USA}
\affiliation{Dept. of Condensed Matter Physics, Weizmann Institute of Science, Rehovot 76100, Israel}

\date{\today}

\begin{abstract}

We observe dynamic nuclear polarization in a GaAs double dot system using two electrons that are
never exchanged with the reservoir. By periodically bringing the system to the mixing point where
the singlet and the triplet $T^+$ states are degenerate, we observe that an excess polarization is built
up. Surprisingly, the pumping procedure is most effective when the total duty cycle equals a multiple
of the Larmor precession time of the Ga and As nuclei. The induced polarization corresponds to
cooling of the underlying nuclear system. The dependence on the dwell time at the mixing point is
found to be non-monotonic.

\end{abstract}

\pacs{Valid PACS appear here}
\maketitle



Single electron spins in a solid state environment constitute a promising candidate for storing and manipulating quantum information \cite{Loss1998}. Recent experiments in few electron quantum dots in GaAs have demonstrated that both spin and charge can be manipulated electrostatically with electron spin coherence times in excess of a few microseconds \cite{Petta2005, Koppens2006}. While the measured spin coherence time $T_2$ is much longer than the time it takes to perform a single quantum operation, the inhomogeneous spin relaxation time $T_2^*$ is short and exceeds the time required for gate operations only by approximately one order of magnitude \cite{Hanson2007}. In GaAs quantum dots it is by now well established that the main source of both the homogeneous and inhomogeneous spin decoherence time is the nuclear spin environment of the host Ga and As atoms \cite{Khaetskii2002, Coish2005, Erlingsson2001, Witzel2006, Taylor2007}. 
Controlling the nuclear spin environment and understanding its dynamics in such systems is therefore a major challenge in the field of solid state quantum computation \cite{Klauser2006, Coish2007, Rudner2007}.

Recently Petta {\it et al.} \cite{Petta2007} showed that dynamic nuclear polarization of the underlying host lattice may be achieved by periodically driving a double dot two electron system from the singlet state $S$ into the triplet state $T^+$ using the hyperfine coupling. Conservation of total angular momentum across the $S$-$T^+$ transition necessitates a spin flip in the nuclear sub system thereby producing a single nuclear spin flip per cycle. Following the transition to the triplet state the two electrons are discarded into a reservoir and a new pair of electrons is reloaded into the singlet state in preparation for the next cycle.  As the sequence is repeated over a large number of cycles, nuclear polarization slightly larger than 1 $\%$ may be achieved.

\begin{figure} [!hbt]
\begin{center}
\includegraphics[width=0.5\textwidth]{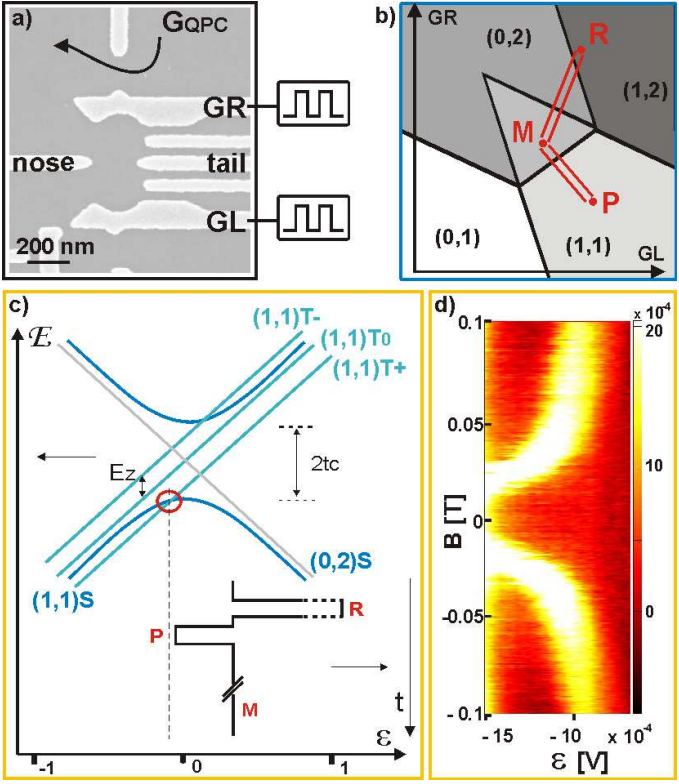}
\end{center}
\caption{a) SEM micrograph of the device. GL and GR control the energy detuning, $\epsilon$, related to the equilibrium occupation of the dot, while gates nose and tail allow to change the interdot tunnel coupling. The average occupation of the dot is sensed by the nearby quantum point contact as a variation of its conductance.  b) Pulse sequence shown within the double dot charge stability diagram. R indicates the reload point, M the measurement point and P the probing point. c) Relevant energy levels at the (0,2)-(1,1) charge transition  as a function of detuning. The three triplet states in the (1,1) configuration are split by the Zeeman energy. The red circle marks the crossing point between singlet and triplet $T^+$. The pulse sequence sketched as a function of detuning and time is used to measure the position of the S-$T^+$ transition as a function of magnetic field. d) Measurement of the position of the $S$-$T^+$ transition as a function of detuning $\epsilon$ and external magnetic field $B_{ext}$. 
\label{Fig1}   }
\end{figure}

In this Letter we present a new scheme for dynamic nuclear polarization that involves only a single pair of electrons. Our pumping cycle works most effectively when the duty cycle of the experiment equals a multiple of the nuclear Larmor precession time. Under these conditions cooling of the nuclear subsystem is observed.

Our measurements are conducted on a GaAs double dot system containing only two electrons. We restrict the phase space to two charge configurations: two electrons in one dot (schematically represented as (0,2)) and one electron in each dot (represented as (1,1)). The double quantum dot (dQD) is created using e-beam patterned PdAu gates that are evaporated on the surface of a GaAs/AlGaAs heterostructure containing a 2 dimensional electron gas (2DEG) situated at 91 nm below the surface with density= $1.3\cdot 10^{15} m^{-2}$ and mobility= $4.9 \cdot 10^6 cm^2/V\cdot s$.
Gate R (GR) and gate L (GL) control the number of electrons in the right and left dot respectively while gates nose and tail allow for an independent control of the tunnel coupling ($t_c$) between the dots (see Fig.~\ref{Fig1}(a)). Both GR and GL are connected via attenuated coaxial cables to a Textronix AWG520 arbitrary waveform generator (with rise time $<$ 1.5 ns) that allows fast pulsing of these gates.

The nearby quantum point contact (QPC) allows the measurement of the double dot charge configurations \cite{Field1993}. We tune the QPC conductance to $\approx0.8 e^2/h$, where it is most sensitive to changes of the surrounding electrostatic potentials. A change of the dot's average occupancy is reflected in a change of the QPC conductance. This is measured in a current bias scheme (5 nA excitation) with standard lock-in technique at a frequency of 137 Hz and 300 ms time constant.

The device is placed in a dilution refrigerator with $T_{base}$ = 10 mK and $T_e$ = 120 mK (established by measuring the width of Coulomb blockade peaks). An external magnetic field of $\lesssim 150$~mT is applied perpendicular to the 2DEG.

Measurements presented here concentrate on the (0,2)-(1,1) charge transition. A particular combination of the voltage applied to gates GL and GR, denoted here by $\epsilon$, allows continuous detuning from one charge configuration to the other. In the absence of spin dependent terms, the ground state in all possible charge configurations is the singlet state. A weak external magnetic field induces a splitting of the three triplet, $T^-$, $T^0$ and $T^+$. While in the (0,2) charge configuration the triplet states are separated from the ground state by a large exchange energy, in the (1,1) configuration the Zeeman energy lowers the $T^+$ state below that of the singlet, as shown in Fig.~\ref{Fig1}(c). A degeneracy point between S and $T^+$, marked by the red circle, is thereby generated at a particular value of $\epsilon$ that depends on the Zeeman energy.

It is important to notice that the total nuclear magnetic field entering the Zeeman energy is $\mathbf{B_{tot}}$=$\mathbf{B_{ext}}$+$\mathbf{B_{nuclei}}$. $\mathbf{B_{nuclei}}$ is a linear combination of the left and right dots nuclear fields, $\mathbf{B_{nuc,R}}$ and  $\mathbf{B_{nuc,L}}$, with a weight that is proportional to the occupation of the two dots.

The measurement of the $S$-$T^+$ degeneracy point is done using the pulse sequence presented in Fig.~\ref{Fig1}(b) and (c). The cycle begins by loading two electrons into the (0,2) singlet state (at reload point {\bf R}). The loading procedure requires typically 200 ns.  Then $\epsilon$ is moved adiabatically with respect to the tunnel coupling ($t_{tunneling}=\hbar/t_c \approx 0.1$ ns) to point {\bf P} where the $S$-$T^+$ transition is probed for $t_P=100$ ns. When point {\bf P} coincides with the $S$~-~$T^+$ degeneracy point a transition from the singlet into the $T^+$ state may occur together with a nuclear spin flip. The typical time for a $S$-$T^+$ transition is approximately $t_{mixing}$=~10~ns~\cite{Koppens2006}. The system is then brought back to the measurement point {\bf M} for the read-out of the final state $|f\rangle$. Since the tunneling between the two charge configurations conserves spin there are two possible outcomes: if $|f\rangle$ is a singlet the transition to (0,2)S is allowed and the QPC detects two electrons in dot R. If $|f\rangle$ is the triplet $T^+$ state, it will be blocked in $(1,1)T^+$ since $(0,2)T^+$ is energetically not allowed and the QPC detects only one electron in dot R. This behavior is coined spin-blockade \cite{Weinmann1994, Ono2002} and occurs in the triangular region in Fig.\ref{Fig1}(b). \emph{Spin to charge} conversion technique has been previously employed \cite{Elzerman2004, Johnson2005}. After the measurement stage the system is reinitialized at point R and the cycle repeats itself.   
The measured signal at the QPC is a time-averaged measurement of the dot occupation throughout the entire cycle. Since $t_M$ $>$ 90 $\%$ of the total pulse time (here $t_M$= 10-20~$\mu$s) we primarily obtain information on point M.

\begin{figure} [!hbt]
\begin{center}
\includegraphics[width=0.5\textwidth]{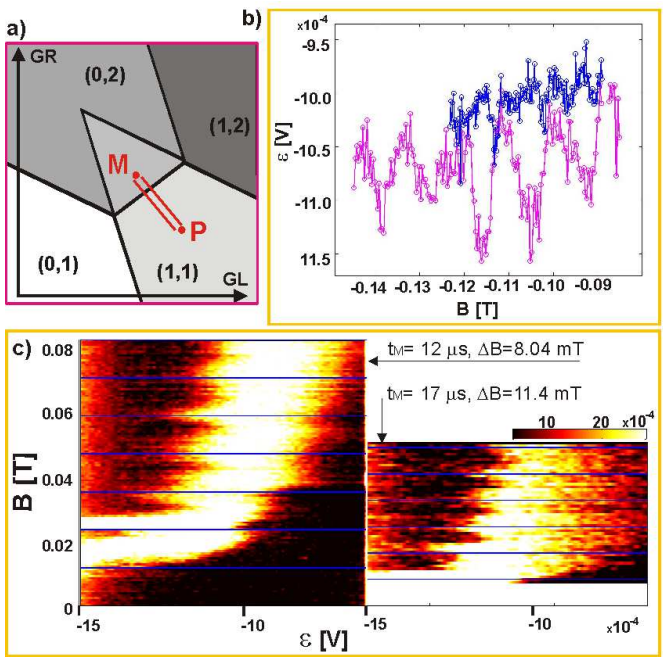}
\end{center}

\caption{a) New pulse sequence used to measure the position of the $S$ to $T^+$ transition when the reload part is left out. In order to probe the S-$T^+$ transition the system is moved to point {\bf P} for a typical time $t_P$=100 ns, then it is brought back to the measurement position {\bf M} for a time $t_M$= 10-20 $\mu$s $>$ 90~$\%$ of the total pulse time. b) Measurement of the position of the S-$T^+$ transition as a function of magnetic field taken with two different pulse sequences: one including a reload part (blue line) and one without reload part (magenta line). c) Measurements of S-$T^+$ transition taken with two different measurement times. The periodicity in magnetic field shows  $\Delta$B= 11.4 mT for $t_M$=12 $\mu$s and $\Delta$B= 8 mT for $t_M$=17 $\mu$s. The blue lines are a guide to the eye helping to compare periodicities. 
\label{Fig2} 
 }
\end{figure}

Fig.\ref{Fig1}(d) shows a measurement of the position of the $S$~-~$T^+$ transition as a function of $\epsilon$ and $B_{ext}$. The plotted QPC conductance reflects the return probability as a singlet. Dark regions indicate that no mixing occurred and the system remained in a singlet state. The bright funnel shaped feature maps the position of the $S$-$T^+$ transition as a function of $B_{ext}$ and $\epsilon$ reflecting a finite probability of being blocked in $T^+$.

What happens if we leave out the reload part from the pulse sequence? Now one must account for the history in order to know the initial state of each cycle. However, given that the relaxation time between the triplet and singlet states (of the order of ms) is shorter than our averaging time (300 ms), when the detuning is away from the $S$-$T^+$ degeneracy point the system will relax to the singlet (0,2) state and remain there \cite{Johnson2005}. Hence, only when the detuning is at the $S$-$T^+$ degeneracy point there will be a finite probability to observe a (1,1) charge state. Therefore a funnel similar to the case with reload should be observed. 

Surprisingly, we record an oscillatory behaviour of the $S$-$T^+$ transition as a function of $B_{ext}$. In Fig.~\ref{Fig2}(b) we plot the values of $B_{ext}$ and $\epsilon$ delimiting the left-most side of the $S$-$T^+$ transition as determined from the QPC conductance plots (see Fig.\ref{Fig1}(d) and Fig.~\ref{Fig2}(c)) in the presence and absence of reload. Whereas the measurement with repeated reload (blue line) shows no magnetic field dependence, in the measurement without reload (magenta line) the position of the $S$-$T^+$ transition dips below the blue line at periodic values in $B_{ext}$. The dips show that the $S$~-~$T^+$ transition shifts towards more negative values of $\epsilon$. From the energy diagram in Fig.\ref{Fig1}(c) we see that this corresponds to a decrease of the Zeeman energy, happening when $\mathbf{B_{tot}}$=$\mathbf{B_{ext}}$+ $\mathbf{B_{HF}}$ decreases. For a fixed $\mathbf{B_{ext}}$, blue points and magenta points are positioned at different values of $\epsilon$, meaning that $\mathbf{B_{HF}}$ changed. The measured change in $\mathbf{B_{HF}}$ is opposite the external magnetic field, thus an accumulation of nuclei with spin up has been created. Such population corresponds to cooling the nuclei. 

The shift of the funnel to more negative detuning can be quantitatively translated to a change in total field using the local slope of the bare funnel (with reload shown in Fig.\ref{Fig1}(d)) at the measurement field. The shift in detuning is translated to approximately 30-50 mT, which corresponds to 1 $\%$ of the total nuclear polarization \cite{Paget1977}.

In order to build up such nuclear polarization we need to repeat the pulse cycle at the $S$-$T^+$ degeneracy point for a few minutes. When measuring the decay time of the created polarization we observe its full decay after about 15 minutes. The pulses are switched off during this time and just briefly turned on at the end in order to measure the actual position of the $S$-$T^+$ transition.

\begin{figure} [!hbt]
\begin{center}
\includegraphics[width=0.5\textwidth]{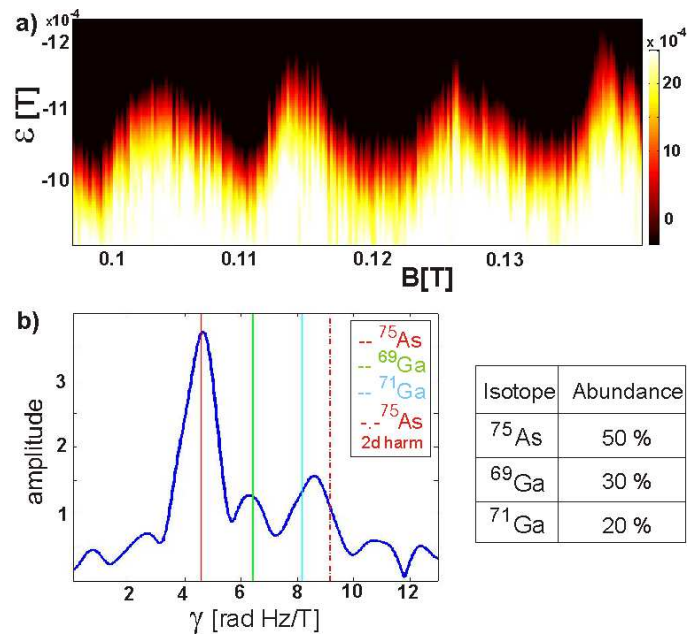}
\end{center}
 
\caption{a) Detailed measurement of the position of the S-$T^+$ transition as a function of $\epsilon$ and $B_{ext}$. b) Fourier transform of detailed measurements. The red, green and blue lines indicate what the tabulated value for the gyromagnetic factor of the three isotopes $^{75}$As,$^{69}$Ga and $^{71}$Ga are \cite{NMRtable}. Since the resolution in Fourier space is low, the peak corresponding to $^{71}$Ga is not clearly distinguishable from the second harmonic of $^{75}$As (dotted red line). The table lists the abundance of each species. 
\label{Fig4}   }
\end{figure}

We also analyze the dependence of the oscillatory behaviour as a function of the measurement time $t_M=10-17\ \mu$s and the probing time $t_P=100-300$~ns, $t_M + t_P$ being the total pulse duration. The two scans in Fig.~\ref{Fig2}(c) show the period of the oscillations in magnetic field $\Delta$B($t_M$) for two different measurement times $t_M=12\ \mu$s and $t_M=17\ \mu$s, while $t_P=200$ ns is kept fixed.  We measure $\Delta B$ for four different values of $t_M$, and find that the gyromagnetic factor $\gamma=2\pi/(\Delta B \cdot t_M)$ is constant and equals $\gamma=4.6 \cdot 10^7$~radHz/T, consistent with the gyromagnetic factor tabulated for $^{75}$As~\cite{NMRtable}. To further confirm this fact, we measured a portion of the funnel with higher trace resolution (see Fig.~\ref{Fig4}(a)): the whole scan of four oscillations takes about 24 hours.

Fourier analysis, Fig.~\ref{Fig4}(b), clearly shows a peak at the Larmor frequency of $^{75}$As. Furthermore, the characteristic frequency of $^{69}$Ga is also visible.  Its other isotope $^{71}$Ga is however hidden by the large second harmonic of $^{75}$As, because of the low resolution in Fourier space. We note that the intensities of the peaks corresponding to the different nuclei is also consistent with their relative abundance inside GaAs \cite{NMRtable}.

\begin{figure} [!hbt]
\begin{center}
\includegraphics[width=0.35\textwidth]{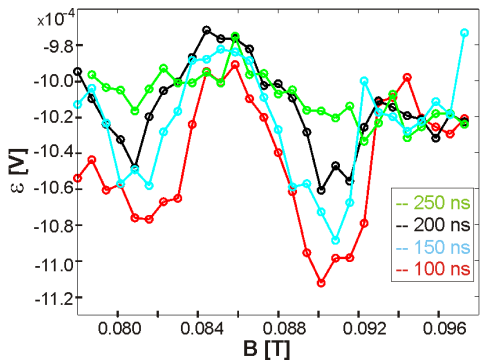}
\end{center}

\caption{ Position of the $S$-$T^+$ transition as a function of $\epsilon$ and $B_{ext}$ measured with different probing times. The amplitude changes, being maximal for 100 ns and decreasing for longer times as well as shorter ones (not shown here).  
\label{Fig3}     }
\end{figure}

Finally, when changing the probing time, while $t_M$ is kept fix to 12~$\mu$s, we observed a change in the amplitude of the oscillations. The maximal amplitude is measured for $t_{P}=100$ ns.  For shorter times the effect vanishes when the probing time is comparable or shorter than the mixing time. For longer times the amplitude decreases and the dips disappear at about $t_P=300$ ns (see Fig.\ref{Fig3}).

In conclusion, we have shown that a periodically driven single pair of electrons in a GaAs double quantum dot through the $S$-$T^+$ transition induces dynamic polarization of the underlying nuclei. The pumping cycle is most efficient when the cycle time corresponds to a multiple of the nuclear Larmor precession time. The gyromagnetic factor of all three nuclear species ($^{75}$As, $^{69}$Ga and $^{71}$Ga) extracted from the measurements confirmed the partial polarization of all three isotopes. 
We have also shown that the probing time directly influences the amplitude of the polarization, with a peak at $t_P=100$ ns. The maximal polarization we have obtained is approximately 1~$\%$ of the total nuclear polarization, and its direction being opposite the external magnetic field corresponds to cooling of the nuclei.

We are grateful for useful discussions with G.~Barak, B.~Verdene, M.~Zaffalon, N.~Akerman, O.~Raslin, \mbox{H.-A.}~Engel, B.~I.~Halperin, J.~J.~Krich, M.~Stopa, L.~S.~Levitov, M.~S.~Rudner, Y.~Gefen, A.~Romito and D.~Loss.  

We acknowledge support from the National Science Foundation 
under PHY-0653336 and the U.S. Army Research Office under W911NF-05-1-0476.

\bibliography{NuclearPolBib}

\end{document}